# Slepian Spatial-Spectral Concentration Problem on the Sphere: Analytical Formulation for Limited Colatitude–Longitude Spatial Region

Alice P. Bates, *Member, IEEE*, Zubair Khalid, *Member, IEEE*, and Rodney A. Kennedy, *Fellow, IEEE*

*Abstract*—In this paper, we develop an analytical formulation for the Slepian spatial-spectral concentration problem on the sphere for a limited colatitude–longitude spatial region on the sphere, defined as the Cartesian product of a range of positive colatitudes and longitudes. The solution of the Slepian problem is a set of functions that are optimally concentrated and orthogonal within a spatial or spectral region. These properties make them useful for applications where measurements are taken within a spatially limited region of the sphere and/or a signal is only to be analyzed within a region of the sphere. To support localized spectral/spatial analysis, and estimation and sparse representation of localized data in these applications, we exploit the expansion of spherical harmonics in the complex exponential basis to develop an analytical formulation for the Slepian concentration problem for a limited colatitude–longitude spatial region. We also extend the analytical formulation for spatial regions that are comprised of a union of rotated limited colatitude–longitude subregions. By exploiting various symmetries of the proposed formulation, we design a computationally efficient algorithm for the implementation of the proposed analytical formulation. Such a reduction in computation time is demonstrated through numerical experiments. We present illustrations of our results with the help of numerical examples and show that the representation of a spatially concentrated signal is indeed sparse in the Slepian basis.

*Index Terms*—Spatial-spectral concentration problem, Slepian functions, 2-sphere (unit sphere), spherical harmonics.

## I. INTRODUCTION

SIGNALS on the sphere appear in a wide range of applications in diverse fields such as geophysics [1], [2], computer graphics [3], [4], cosmology [5]–[7], medical imaging [8], [9], electron microscopy [10] and acoustics [11], [12]. Spherical harmonics are the archetype set of complete, orthonormal functions on the sphere. However, spherical harmonics are a global basis so they do not efficiently represent a signal in a restricted region of the sphere; they are also not orthogonal except for when the region is the whole sphere.

The solution of the Slepian spatial-spectral concentration problem on the sphere are the Slepian functions which form an alternative complete basis that is not only orthogonal on the sphere, but also orthogonal within a given region on the sphere. Slepian functions are also optimally concentrated within the region of the sphere on which they are defined [1], [13]. Consequently, Slepian functions on the sphere have been used for localized spectral and spatial analysis [14]–[16], and signal estimation from incomplete measurements [17], [18], and sparse and efficient representations of spherical in signals in a wide range of applications found in geophysics [16], [19], [20], cosmology and planetary studies [21], [22], optics [18] and computer graphics [23], to name a few.

Slepian functions on the sphere arise as the solution to the problem, first considered in one-dimension by Slepian, Pollak and Landau [24]–[27], of finding functions that are band-limited and maximally concentrated within a closed region on the sphere (or spatially limited and optimally concentrated within some band-limit). For an arbitrary region, there is no closed-form solution to this problem and the Slepian functions for a given region are calculated numerically. However analytical expressions are desirable as they allow exact computation and the development of computationally efficient algorithms. Simons and Dahlen developed an analytical expression for computing Slepian functions concentrated in a polar cap or a polar gap region on the sphere [28]. The polar gap region is useful in geophysics; it appears in satellite data of the gravitational or magnetic field potential of the earth where the pair of axisymmetric polar caps do not have data coverage due to the inclined orbits of the satellite.

Another useful regions on the sphere is the limited colatitude-longitude region, defined as a Cartesian product of a range of colatitudes and longitudes. For example, limited colatitude-longitude regions appear in the following applications: the cosmic microwave background radiation observed from earth is approximately seen within a limited colatitude-longitude region [29], [30], often a signal of interest in geophysics such as magnetic or gravitational potential are considered between lines of co-latitude and longitude [31], the projection of a rectangular sound source in acoustics or light source in optics on the sphere forms a limited colatitude-longitude region on the sphere [32] and a limited colatitude-longitude region is used to

Manuscript received August 9, 2016; revised December 8, 2016; accepted December 14, 2016. Date of publication December 30, 2016; date of current version January 20, 2017. The associate editor coordinating the review of this manuscript and approving it for publication was Dr. Yuichi Tanaka. This work was supported by the Australian Research Council's Discovery Projects funding scheme under Project DP150101011.
A. P. Bates and R. A. Kennedy are with the Research School of Engineering, The Australian National University, Canberra, ACT 0200, Australia (e-mail: alice.bates@anu.edu.au; rodney.kennedy@anu.edu.au).
Z. Khalid is with the Department of Electrical Engineering, School of Science and Engineering, Lahore University of Management Sciences, Lahore 54792, Pakistan (e-mail: zubair.khalid@lums.edu.pk).
Color versions of one or more of the figures in this paper are available online at http://ieeexplore.ieee.org.
Digital Object Identifier 10.1109/TSP.2016.2646668

describe possible angles of arrival in communications [33], [34]. As the limited colatitude-longitude region is widely applicable, it would be useful to have an analytical formulation for solving the Slepian problem in this region.

In this work, we develop an analytical formulation for the Slepian spatial-spectral concentration problem on the sphere for a limited colatitude-longitude spatial region on the sphere by exploiting the expansion of spherical harmonics in the complex exponential basis. We also extend this analytical formulation to enable the computation of Slepian functions concentrated within an arbitrary region of the sphere comprised of a union of rotated limited colatitude-longitude subregions. By exploiting various symmetries of the proposed formulation, we develop a computationally efficient algorithm for computing Slepian functions over a limited colatitude-longitude region using the proposed formulation. We demonstrate the reduction in computation time through numerical experiments. We use further numerical experiments to illustrate our results and show that the representation of a spatially concentrated signal is indeed sparse in the Slepian basis.

The remainder of the paper is organized as follows. We present the necessary mathematical background for signals on the sphere and their representation in the spherical harmonic domain in Section II before reviewing Slepian functions on the sphere. We develop an analytical formulation for the Slepian spatio-spectral concentrated problem on the sphere for a limited colatitude-longitude region on the sphere in Section III. We then extend this analytical formulation for an arbitrary region on the sphere. In Section III we also present the properties of Slepian functions concentrated in a limited colatitude-longitude region and illustrate their use with examples. We then develop a computationally efficient algorithm for implementation of the analytical formulation in Section IV and carry out computational complexity analysis of the proposed algorithm. Concluding remarks are then made in Section V.

## II. MATHEMATICAL PRELIMINARIES

In order to clarify the notation adopted throughout the paper, we present the relevant mathematical background for signals defined on the sphere, their spectral domain representation and the rotation of signals on the sphere. We also briefly review Slepian spatial-spectral concentration problem on the sphere.

### A. Signals on the Sphere

The spherical domain, also referred as sphere or 2-sphere or unit sphere, is denoted by $\mathbb{S}^2$ and is defined as $\mathbb{S}^2 \triangleq \{\boldsymbol{x} \in \mathbb{R}^3 : |\boldsymbol{x}| = 1\} \subset \mathbb{R}^3$, where $|\cdot|$ represents Euclidean norm [35]. A point on $\mathbb{S}^2$ is given by a unit vector $\hat{\boldsymbol{x}} \equiv \hat{\boldsymbol{x}}(\theta, \phi) \triangleq (\sin\theta\cos\phi, \sin\theta\sin\phi, \cos\theta)' \in \mathbb{R}^3$, where $(\cdot)'$ denotes the vector transpose operation, $\theta \in [0, \pi]$ is the colatitude that is measured with respect to the positive $z-$ axis and $\phi \in [0, 2\pi)$ is the longitude which is measured with respect to the positive $x-$ axis in the $x - y$ plane.

We consider the complex-valued square-integrable functions defined on the sphere. The set of such functions form a Hilbert space denoted by $L^2(\mathbb{S}^2)$ equipped with the inner product given by [35]

$$\langle f, h \rangle \triangleq \int_{\mathbb{S}^2} f(\hat{\boldsymbol{x}})\overline{h(\hat{\boldsymbol{x}})}\, ds(\hat{\boldsymbol{x}}), \tag{1}$$

for two functions $f$ and $h$ defined on $\mathbb{S}^2$. Here $ds(\hat{\boldsymbol{x}}) = \sin\theta\, d\theta\, d\phi$ is the differential area element on $\mathbb{S}^2$. The inner product induces a norm $\|f\| \triangleq \langle f, f \rangle^{1/2}$. We refer the functions with finite energy (finite induced norm) as signals on the sphere. We also define $\langle f, g \rangle_R \triangleq \int_R f(\hat{\boldsymbol{x}})\overline{g(\hat{\boldsymbol{x}})}\, ds(\hat{\boldsymbol{x}})$ as the inner product on the region $R$ and $\|f\|_R^2 \triangleq \langle f, f \rangle_R$ as the energy of the signal $f$ in $R$.

### B. Spherical Harmonic Domain Representation

The spherical harmonic function $Y_\ell^m(\theta, \phi)$ for integer degree $\ell \geq 0$ and integer order $|m| \leq \ell$ is defined as [35], [36]

$$Y_\ell^m(\hat{\boldsymbol{x}}) = Y_\ell^m(\theta, \phi) \triangleq \sqrt{\frac{2\ell+1}{4\pi}\frac{(\ell-m)!}{(\ell+m)!}}\, P_\ell^m(\cos\theta)e^{im\phi}, \tag{2}$$

where $P_\ell^m$ denotes the associated Legendre function of integer degree $\ell$ and integer order $m$ and is defined as [35]

$$P_\ell^m(x) = \frac{(-1)^m}{2^\ell \ell!}(1-x^2)^{m/2}\frac{d^{\ell+m}}{dx^{\ell+m}}(x^2-1)^\ell$$

$$P_\ell^{-m}(x) = (-1)^m \frac{(\ell-m)!}{(\ell+m)!}P_\ell^m(x),$$

for $|x| \leq 1$ and $m \geq 0$. Spherical harmonics functions (spherical harmonics for short) are orthonormal over the sphere with $\langle Y_\ell^m, Y_p^q \rangle = \delta_{\ell,p}\delta_{m,q}$, where $\delta_{m,q}$ is the Kronecker delta function: $\delta_{m,q} = 1$ for $m = q$ and is zero otherwise. Spherical harmonics form a complete orthonormal set of basis functions for $L^2(\mathbb{S}^2)$ [35], and therefore we can expand any signal $f \in L^2(\mathbb{S}^2)$ as

$$f(\hat{\boldsymbol{x}}) = \sum_{\ell,m}^{\infty}(f)_\ell^m Y_\ell^m(\hat{\boldsymbol{x}}), \tag{3}$$

where $\sum_{\ell,m}^{\infty} \triangleq \sum_{\ell=0}^{\infty}\sum_{m=-\ell}^{\ell}$, that is, we have expressed the double summation as a single summation for notational convenience and

$$(f)_\ell^m \triangleq \langle f, Y_\ell^m \rangle = \int_{\mathbb{S}^2} f(\hat{\boldsymbol{x}})\overline{Y_\ell^m(\hat{\boldsymbol{x}})}\, ds(\hat{\boldsymbol{x}}) \tag{4}$$

denotes the spherical harmonic coefficient of degree $\ell$ and order $m$ which form the spectral (spherical harmonic) domain representation of a signal.

The signal $f \in L^2(\mathbb{S}^2)$ is defined to be band-limited at degree $L$ if $(f)_\ell^m = 0$ for $\ell \geq L$. The set of band-limited signals forms an $L^2$ dimensional subspace of $L^2(\mathbb{S}^2)$, which is denoted by $\mathcal{H}_L$. For the spectral domain representation of a band-limited signal $f \in \mathcal{H}_L$, we define the column vector containing spherical harmonic coefficients as $\mathbf{f} \triangleq \left((f)_0^0, (f)_1^{-1}, (f)_1^0, (f)_1^1, (f)_2^{-2}, \cdots, (f)_{L-1}^{L-1}\right)'$ of size $L^2$.

### C. Rotation on the Sphere

The rotation operator $\mathcal{D}(\varphi, \vartheta, \omega)$ rotates a function on the sphere by an angle $\omega$ around the $z$-axis, followed by an angle $\vartheta$ around the $y$-axis and finally an angle $\varphi$ around the $z$-axis, where we use the $zyz$ rotation convention, and the axis and rotations follow a right-handed convention [35]. Applying the rotation operator to a function is realised by inverse rotation of the coordinate system with

$$(\mathcal{D}(\varphi, \vartheta, \omega)f)(\hat{\boldsymbol{x}}) = f(\mathbf{R}^{-1}\hat{\boldsymbol{x}}), \quad (5)$$

where $\mathbf{R}$ is the $3 \times 3$ rotation matrix corresponding to the rotation operator $\mathcal{D}(\varphi, \vartheta, \omega)$ [35]. $\mathcal{D}(\varphi, \vartheta, \omega)^{-1}$ denotes the inverse of the rotation operator and is given by $\mathcal{D}(\varphi, \vartheta, \omega)^{-1} = \mathcal{D}(\pi - \omega, \vartheta, \pi - \varphi)$.

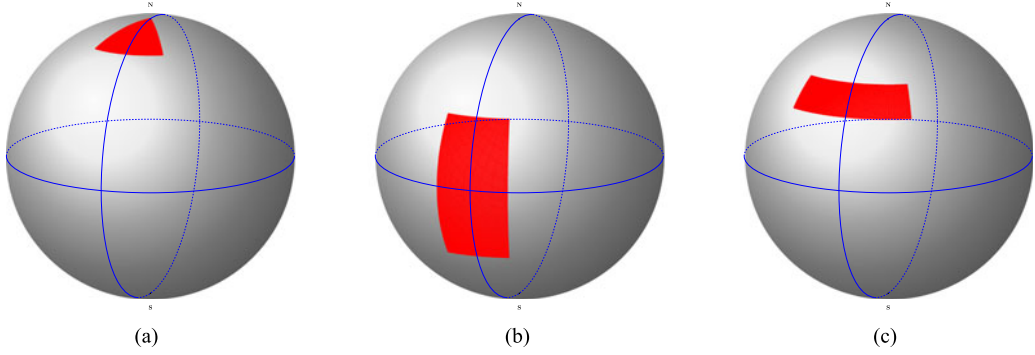

Fig. 1. The region $\tilde{R}$ on the sphere, defined in (13), is shaded in red for different parameters: (a) $[\theta_1, \theta_2] = [0, \pi/6]$, $[\phi_1, \phi_2] = [-\pi/6, \pi/6]$, (b) $[\theta_1, \theta_2] = [\pi/3, 2\pi/3]$, $[\phi_1, \phi_2] = [-\pi/12, \pi/12]$, and (c) $[\theta_1, \theta_2] = [\pi/4, \pi/3]$, $[\phi_1, \phi_2] = [-\pi/6, \pi/6]$. $\phi = 0$ iso-longitude and $\theta = \pi/2$ iso-colatitude (equator) lines are shown in blue. As the angle $\phi$ is periodic in $2\pi$, $-\phi$ is equal to $2\pi - \phi$.

The spherical harmonic coefficient of the rotated output signal of degree $\ell$ and order $m$ is a linear combination of different order spherical harmonic coefficients of the same degree of the original function with

$$\left(\mathcal{D}(\varphi, \vartheta, \omega)f\right)_\ell^m \triangleq \langle D(\varphi, \vartheta, \omega)f, Y_\ell^m \rangle$$
$$= \sum_{m'=-\ell}^{\ell} D_\ell^{m,m'}(\varphi, \vartheta, \omega)(f)_\ell^{m'}, \quad (6)$$

where $D_\ell^{m,m'}(\varphi, \vartheta, \omega)$ is the Wigner-$D$ function given by

$$D_\ell^{m,m'}(\varphi, \vartheta, \omega) = e^{-im\varphi} d_\ell^{m,m'}(\vartheta) e^{-im'\omega}, \quad (7)$$

and $d_\ell^{m,m'}(\vartheta)$ is the Wigner-$d$ function [35].

### D. Spatial-Spectral (Slepian) Concentration Problem on the Sphere

The Slepian spatial-spectral concentration problem [24]–[27] on the sphere for finding functions that are band-limited (or space-limited) with maximal energy concentrated in the given spatial (or spectral) region has been extensively investigated [1], [13], [14], [37]. In order to maximize the spatial (energy) concentration of a unit-energy band-limited signal $h \in \mathcal{H}_L$ within the spatial region $R \subset \mathbb{S}^2$, we seek to maximize the spatial concentration ratio $\lambda$ given by [13],

$$\lambda = \frac{\|h\|_R^2}{\|h\|^2}, \quad 0 \leq \lambda < 1, \quad (8)$$

which can be equivalently expressed in spectral domain as

$$\lambda = \frac{\sum_{\ell,m}^{L-1} \sum_{p,q}^{L-1} \overline{(h)_\ell^m} (h)_p^q K_{\ell m, pq}}{\sum_{\ell,m}^{L-1} \overline{(h)_\ell^m} (h)_\ell^m}, \quad (9)$$

where

$$K_{\ell m, pq} \triangleq \int_R Y_p^q(\hat{\boldsymbol{x}}) \overline{Y_\ell^m(\hat{\boldsymbol{x}})} ds(\hat{\boldsymbol{x}}). \quad (10)$$

The problem to maximize the concentration ratio in (9) can be solved as an algebraic eigenvalue problem [13]

$$\sum_{\ell=0}^{L-1} \sum_{m=-\ell}^{\ell} K_{\ell m, pq} (h)_p^q = \lambda (h)_\ell^m, \quad (11)$$

which can be written in matrix form as

$$\mathbf{Kh} = \lambda \mathbf{h}, \quad (12)$$

where the matrix $\mathbf{K}$ has dimension $L^2 \times L^2$ and contains elements $K_{\ell m, pq}$, given in (10), with similar indexing adopted for $\mathbf{h}$.

## III. SPATIAL-SPECTRAL CONCENTRATION PROBLEM FOR LIMITED COLATITUDE-LONGITUDE SPATIAL REGION

In order to solve the spatial-spectral concentration problem (11), we are first required to evaluate the matrix $\mathbf{K}$. Since there are no quadrature rules for evaluating the integral over the region $R$ in general, $K_{\ell m, pq}$, given in (10), must be computed numerically. Analytic expressions have been devised in the literature to compute $K_{\ell m, pq}$ for the azimuthally symmetric (polar cap)[1] and polar gap regions [28].

Here, we revisit the spatial-spectral concentration problem on the sphere for limited colatitude-longitude spatial region $\tilde{R}$ defined as a Cartesian product of a range of limited colatitudes and limited longitudes, that is,

$$\tilde{R} \triangleq \{(\theta, \phi) : \theta_1 \leq \theta \leq \theta_2, \phi_1 \leq \phi \leq \phi_2\}. \quad (13)$$

We note that the region $\tilde{R}$ is parameterized by four parameters: $\theta_1, \theta_2, \phi_1$ and $\phi_2$. For example, the region $\tilde{R}$ is shown in Fig. 1 for different values of $\theta_1, \theta_2, \phi_1$ and $\phi_2$, where it can be observed that a different choice of parameters gives rise to the regions of different shapes on the sphere.

---

[1]Since any rotationally symmetric region, that is symmetric with respect to the rotation around its axis, can be represented as an azimuthally symmetric region by appropriately rotating the region [35], $K_{\ell m, pq}$ can be computed analytically for any rotationally symmetric region.

For the limited colatitude-longitude spatial region $\tilde{R}$, we derive an analytic expression to compute $K_{\ell m, pq}$, given in (10) which consequently, enables the accurate computation of band-limited functions with optimal concentration in the spatial region $\tilde{R}$. It is expected that the proposed development would support the signal analysis in applications [29]–[34] where the signals/data-sets are measured/concentrated over the limited colatitude-longitude spatial region $\tilde{R}$.

### A. Computation of Matrix $\mathbf{K}$

*Theorem 1:* For a limited colatitude-longitude spatial region $\tilde{R}$ defined in (13), the elements of the matrix $\mathbf{K}$ given in (10), have the following analytical expression

$$K_{\ell m, pq} = \sum_{m'=-\ell}^{\ell} \sum_{q'=-p}^{p} F_{m',m}^{\ell} F_{q',q}^{p} Q(m'+q') S(q-m), \quad (14)$$

where

$$Q(m) = \begin{cases} \frac{1}{4}\left(2im(\theta_2 - \theta_1) + e^{2im\theta_1} - e^{2im\theta_2}\right), & |m| = 1 \\ \frac{1}{m^2-1}\left(e^{im\theta_1}(-\cos\theta_1 + im\sin\theta_1) \right. \\ \left. + e^{im\theta_2}(\cos\theta_2 - im\sin\theta_2)\right), & |m| \neq 1, \end{cases} \quad (15)$$

$$S(m) = \begin{cases} \phi_2 - \phi_1, & m = 0 \\ \frac{i}{m}\left(e^{im\phi_1} - e^{im\phi_2}\right), & m \neq 0, \end{cases} \quad (16)$$

and

$$F_{m',m}^{\ell} = (-i)^m \sqrt{\frac{2\ell+1}{4\pi}} \Delta_{m',m}^{\ell} \Delta_{m',0}^{\ell}, \quad (17)$$

where

$$\Delta_{m,n}^{\ell} \triangleq d_{m,n}^{\ell}(\pi/2).$$

*Proof:* In order to determine an analytic expression for the computation of the matrix elements $K_{\ell m, pq}$, defined in (10), for a limited colatitude-longitude region $\tilde{R}$, we first note the following relation for associated Legendre polynomials

$$P_{\ell}^{m}(\cos\theta) = \sqrt{\frac{(\ell+m)!}{(\ell-m)!}} d_{m,0}^{\ell}(\theta), \quad (18)$$

where $d_{m,n}^{\ell}(\cdot)$ denotes the Wigner-$d$ function of degree $\ell$ and orders $m, n$ and has the following expansion in terms of complex exponentials [35], [38]

$$d_{m,n}^{\ell}(\theta) = i^{n-m} \sum_{m'=-\ell}^{\ell} \Delta_{m',m}^{\ell} \Delta_{m',n}^{\ell} e^{im'\theta}. \quad (19)$$

Using (18) and (19), we write $K_{\ell m, pq}$, given in (10) as

$$\begin{aligned} K_{\ell m, pq} &= \frac{\sqrt{(2\ell+1)(2p+1)}}{4\pi} \\ &\quad \times \int_{\tilde{R}} d_{m,0}^{\ell}(\theta) d_{q,0}^{p}(\theta) e^{i(q-m)\phi} \sin\theta\, d\theta\, d\phi \\ &= \sum_{m'=-\ell}^{\ell} \sum_{q'=-p}^{p} F_{m',m}^{\ell} F_{q',q}^{p} \\ &\quad \times \underbrace{\int_{\theta=\theta_1}^{\theta_2} e^{i(m'+q')\theta} \sin\theta\, d\theta}_{Q(m'+q')} \underbrace{\int_{\phi=\phi_1}^{\phi_2} e^{i(q-m)\phi}\, d\phi}_{S(q-m)}, \end{aligned} \quad (20)$$

which is equivalent to (14). ∎

*Remark 1 (Fast Computation of $\mathbf{K}$):* Using the analytic expression given in Theorem 1, the matrix $\mathbf{K}$ can be computed exactly. We later show the symmetry relations that hold for matrix elements $K_{\ell m, pq}$ and can be exploited to speed-up the computation of $\mathbf{K}$. We elaborate on this when we discuss the fast computation of the matrix $\mathbf{K}$ later in the paper.

### B. Spatial-Spectral Concentration Problem - Analysis

Once the matrix $\mathbf{K}$ is computed exactly using the analytic expression given in Theorem 1, the concentration problem can be solved using its formulation as an algebraic eigenvalue problem given in (12), the solution of which gives $L^2$ band-limited eigenvectors. Each eigenvector represents the spectral domain representation (spherical harmonic coefficients) of the band-limited eigenfunction (in spatial domain) and the eigenvalue associated with each eigenvector represents the concentration of the associated eigenfunction in the spatial region $\tilde{R}$. Since $\mathbf{K}$ is complex-valued and Hermitian symmetric, by definition, the eigenvalues are real and the eigenvectors are orthogonal, we choose them to be orthonormal. Furthermore, the eigenvalues are non-negative as $\mathbf{K}$ is positive-semidefinite which follows from the numerator in (9) that represents the energy of the band-limited function in some spatial region.

Let the eigenvectors of $\mathbf{K}$ and the corresponding eigenfunctions be denoted by $\mathbf{h}_\alpha$ and $h_\alpha(\theta, \phi)$ for $\alpha = 1, 2, \ldots, L^2$, where we index the eigenfunctions such that $0 \leq \lambda_{\alpha+1} \leq \lambda_\alpha < 1$, $\alpha = 1, 2, \ldots, L^2$. With this indexing, the eigenfunction $h_1(\theta, \phi)$ is most concentrated in $\tilde{R}$, while $h_{L^2}(\theta, \phi)$ is most concentrated in $\mathbb{S}^2 \setminus \tilde{R}$.

*1) Orthogonality of Eigenfunctions:* The eigenvectors, by definition, are orthonormal, that is,

$$\mathbf{h}_\alpha^H \mathbf{h}_\beta = \sum_{\ell, m}^{L-1} \overline{(h_\alpha)_\ell^m} (h_\beta)_\ell^m = \delta_{\alpha, \beta}, \quad (21)$$

$$\mathbf{h}_\alpha^H \mathbf{K} \mathbf{h}_\beta = \sum_{\ell, m}^{L-1} \sum_{p, q}^{L-1} \overline{(h_\alpha)_\ell^m} (h_\beta)_p^q K_{pq, \ell m} = \lambda_\alpha \delta_{\alpha, \beta}, \quad (22)$$

where $(\cdot)^H$ denotes the Hermitian of a vector or matrix, which can be equivalently expressed in terms of the associated

eigenfunctions as

$$\|h\|_2^2 = \int_{\mathbb{S}^2} h_\alpha(\hat{\boldsymbol{x}})\overline{h_\beta(\hat{\boldsymbol{x}})}\,ds(\hat{\boldsymbol{x}}) = \delta_{\alpha,\beta}, \quad (23)$$

$$\|h\|_{\tilde{R}}^2 = \int_{\tilde{R}} h_\alpha(\hat{\boldsymbol{x}})\overline{h_\beta(\hat{\boldsymbol{x}})}\,ds(\hat{\boldsymbol{x}}) = \lambda_\alpha\,\delta_{\alpha,\beta}. \quad (24)$$

These relations indicate that the eigenfunctions are not only orthonormal over the sphere but are orthogonal over the spatial region $\tilde{R}$. This double orthogonality is one of the important feature of these eigenfunctions which makes them useful in the analysis of the signal over the spatial region $\tilde{R}$ [14], [17], [21], [39]. We note that the properties of eigenfunctions given in (21)–(24) hold for arbitrary spatial regions [13].

*2) Number of Concentrated Eigenfunctions:* We also note that the sum of the eigenvalues of $\mathbf{K}$ for the spatial region $\tilde{R}$ is given by the trace of $\mathbf{K}$ [13], [35] with,

$$N = \sum_{\alpha=1}^{L^2} \lambda_\alpha = \sum_{\ell,m}^{L-1} K_{\ell m,\ell m}$$
$$= \frac{L^2}{4\pi}\int_{\tilde{R}}\sin\theta d\theta d\phi = \frac{L^2}{4\pi}(\phi_2-\phi_1)(\cos\theta_1 - \cos\theta_2). \quad (25)$$

If the spectrum of eigenvalues has a narrow transition from significant (near unity) to insignificant (near zero) eigenvalues, the sum of the eigenvalues, given by $N$, well-approximates the number of significant eigenvalues.

*3) Slepian Basis:* Since we obtain a set of $L^2$ band-limited orthonormal eigenfunctions as a solution of the (Slepian) spatial-spectral concentration problem, these eigenfunctions span the $L^2$ dimensional subspace $\mathcal{H}_L$ and therefore serve as a complete basis, referred to as the *Slepian basis* [13], for the representation of any band-limited signal. Any band-limited signal $f \in \mathcal{H}_L$ can be expressed in the Slepian basis as

$$f(\hat{\boldsymbol{x}}) = \sum_{\alpha=1}^{L^2} (f)_\alpha\,h_\alpha(\hat{\boldsymbol{x}}), \quad (26)$$

where

$$(f)_\alpha \triangleq \langle f, h_\alpha \rangle = \int_{\mathbb{S}^2} f(\hat{\boldsymbol{x}})\,\overline{h_\alpha(\hat{\boldsymbol{x}})}ds(\hat{\boldsymbol{x}}), \quad (27)$$

denotes the Slepian coefficient of index $\alpha$. Since Slepian functions are orthogonal over the spatial region $\tilde{R}$, the Slepian coefficient can also be determined as

$$(f)_\alpha = \frac{1}{\lambda_\alpha}\int_{\tilde{R}} f(\hat{\boldsymbol{x}})\overline{h_\alpha(\hat{\boldsymbol{x}})}ds(\hat{\boldsymbol{x}}). \quad (28)$$

The signal $f(\hat{\boldsymbol{x}})$ within the spatial region $\tilde{R}$ can be well-approximated by excluding the basis functions with almost zero concentration within the region in the expansion of the signal given in (26), that is, the summation in (26) can be truncated at $J$ such that $\lambda_{J+1} \approx 0$ as

$$f(\hat{\boldsymbol{x}}) \approx \sum_{\alpha=1}^{J} (f)_\alpha\,h_\alpha(\hat{\boldsymbol{x}}), \quad \hat{\boldsymbol{x}} \in \tilde{R}. \quad (29)$$

The quality of approximation of the signal given in (29) within the spatial region $\tilde{R}$ can be measured by defining the quality measure as a ratio of the energy concentration of the approximate representation to the energy of the exact representation within the spatial region, that is,

$$Q(J) = \frac{\int_{\tilde{R}}\left|\sum_{\alpha=1}^{J}(f)_\alpha h_\alpha(\hat{\boldsymbol{x}})\right|^2 ds(\hat{\boldsymbol{x}})}{\int_{\tilde{R}}\left|\sum_{\alpha=1}^{L^2}(f)_\alpha h_\alpha(\hat{\boldsymbol{x}})\right|^2 ds(\hat{\boldsymbol{x}})}$$
$$= \frac{\sum_{\alpha=1}^{J}\lambda_\alpha\left|(f)_\alpha\right|^2}{\sum_{\alpha=1}^{L^2}\lambda_\alpha\left|(f)_\alpha\right|^2}, \quad (30)$$

where we have used the orthogonality of Slepian basis over the spatial region $\tilde{R}$, given in (24), in obtaining the second equality.

*Remark 2 (Truncation at $N$):* Since the number of Slepian basis that are well concentrated in the region is approximately represented by sum of the eigenvalues, $N$, given in (25), the truncation level in (29) can be chosen as $J = N$. We note that such truncation at $J = N$ is based on the assumption that the eigenvalue spectrum has sharp transition from 1 to 0. If for some cases this assumption is not fairly supported, $N$ can be used to estimate the truncation level $J > N$ such that $\lambda_{J+1} \approx 0$.

The representation of the signal within the region $\tilde{R}$ using $N$ basis functions and the computation of Slepian coefficients as an integral over different spatial regions have also been adopted and studied for multi-dimensional Euclidean domains and various geometries [13], [14], [17], [40], [41]. We expect that the accurate computation of the Slepian basis using the proposed formulation for the limited colatitude-longitude region is of great use in applications where the signals are concentrated within some spatial region $\tilde{R}$ [33] or measurements can only be taken over a spatially limited region $\tilde{R}$ [23], [29].

### C. Arbitrary Region of Interest

For the limited colatitude-longitude region $\tilde{R}$, we noted earlier that the different choices of parameters of the region give rise to the regions of different shapes on the sphere (see Fig. 1). This characteristic of the limited colatitude-longitude region $\tilde{R}$ can be exploited to compute the Slepian basis for any arbitrary shaped region. We assume that an arbitrary shaped region $R$ can be partitioned into $M$ disjoint subregions $R_i \cap R_j = \emptyset$, $i \neq j$, as $R = R_1 \cup R_2 \cup \ldots \cup R_M$, where each $R_n$ denotes the limited colatitude-longitude region $\tilde{R}_n$ rotated by an angle $\omega_n$ around the $z$-axis, followed by an angle $\vartheta_n$ around the $y$-axis and finally by an angle $\varphi_n$ around the $z$-axis using the rotation operator $\mathcal{D}(\varphi_n, \vartheta_n, \omega_n)$, as described in Section II-C. We note that each limited colatitude-longitude region $\tilde{R}_n$, $n = 1, 2, \ldots, M$ may have different parameters.

*Corollary 1:* For a given band-limit $L$ and an arbitrary region $R$, let the matrix for arbitrary region be denoted by $\mathbf{K}$ with elements $K_{\ell m, pq}$ given by (10) which can be computed by incorporating the effect of rotation in the harmonic domain, given by (6) and using the partition of the region $R$ into disjoint

subregions, as

$$K_{\ell m,pq} = \sum_{n=1}^{M} \left( \sum_{t=-\ell}^{\ell} \overline{D_{\ell}^{t,m}(\pi - \omega_n, \vartheta_n, \pi - \varphi_n)} \right.$$
$$\left. \times \sum_{r=-p}^{p} D_{p}^{r,q}(\pi - \omega_n, \vartheta_n, \pi - \varphi_n) K_{\ell t,pr}^{n} \right), \quad (31)$$

where $K_{\ell t,pr}^{n}$ denotes the matrix elements, given by Theorem 1, for the limited colatitude-longitude region $\tilde{R}_n$ and depends on the parameters of the region $[\theta_1, \theta_2]$, $[\phi_1, \phi_2]$.

Once $\mathbf{K}$ is computed using (31), the eigenvalue decomposition of $\mathbf{K}$ yields the Slepian basis for the region for a given band-limit $L$. We note that the equivalence between (10) and the formulation in (31) depends on the partition of the region $R$ into $M$ number of rotated limited colatitude-longitude regions. The chosen partitioning of $R$ using an optimal tiling of rotated limited colatitude-longitude regions is the field of finite-element analysis [42], [43] and is beyond the scope of the current work.

### D. Illustration

In Section III we present an analytic formulation for solving the Slepian spatio-spectral problem for a limited colatitude-longitude region $\tilde{R}$ and the properties of Slepian functions in $\tilde{R}$. We here present examples to illustrate the use of Slepian functions in $\tilde{R}$ and demonstrate their properties, using the analytical formulation in Theorem 1 to calculate the matrix $\mathbf{K}$.

*1) Slepian Functions and Eigenvalue Spectrum:* We show the eigenvalue spectrum and Slepian functions band-limited at $L = 25$ for the two limited colatitude-longitude regions, $\tilde{R}$ shown in Fig. 1(a) and Fig. 1(b), which we refer to as Example A and Example B respectively. Fig. 2(a) and (b) show the first 60 eigenvalues in the eigenvalue spectrum for Example A and Example B respectively. The trace of the matrix $\mathbf{K}$ given by (25), which approximates the number of well-concentrated eigenfunctions in the region, is shown by the black dashed line in Fig. 2. As Example A has a smaller area than Example B, it has a smaller number of well-concentrated eigenfunctions with $N = 7$, whereas Example B has $N = 26$.

Fig. 3 shows the magnitude of twelve Slepian functions on the sphere $|h_\alpha(\hat{\boldsymbol{x}})|$, $\alpha = 1, 2, \ldots, 12$ that have the highest concentration for the region $\tilde{R}$ in Example A. Fig. 4 shows the magnitude of the Slepian functions $|h_\alpha(\hat{\boldsymbol{x}})|$, $\alpha = 1, 2, 3, 4, 21, 22, 23, 24, 31, 32, 33$ and $34$ that are well-concentrated in the region $\tilde{R}$ in Example B.

*2) Slepian Basis:* We here present an example to illustrate that the representation of a spatially concentrated band-limited signal in the Slepian basis is sparse and allows for accurate reconstruction when the basis is truncated at $J = N$ using (29). We use a test signal $f(\hat{\boldsymbol{x}})$ obtained from a dark matter distribution of the Universe simulation observed over a partial field of view. The test signal is extracted from the full-sky Horizon Simulation, a simulation derived from the 3-year Wilkinson Microwave Anisotropy Probe (WMAP) observations, at radius $r = 20$ where the radius has units in Mpc. The partial field of view is over a limited colatitude-longitude region approximating the Sloan Digital Sky Survey (SDSS) DR7 [2] quasar binary mask, the mask and the outline of the limited colatitude-longitude region surrounding the mask are shown in Fig. 5. The masked signal is band-limited at spherical harmonic degree $L = 50$ to obtain the spatially concentrated band-limited test signal $f(\hat{\boldsymbol{x}})$ shown in Fig. 6(a).

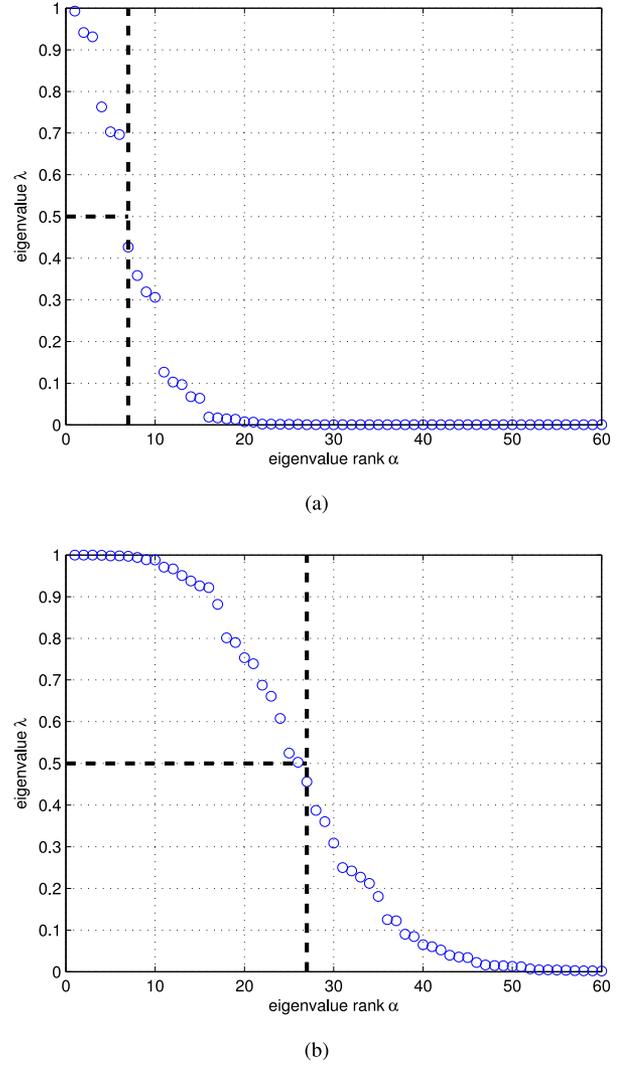

Fig. 2. Eigenvalue spectrum $\lambda_\alpha$, $\alpha = 1, 2, \ldots, 60$ for the Slepian functions with band-limit $L = 25$ concentrated in the limited colatitude-longitude regions shown in (a) Fig. 1(a) with $N = 7$ and (b) Fig. 1(b) with $N = 26$. The number of well-concentrated eigenfunctions is well approximated by $N$ which is shown by the dashed black line.

Since Slepian functions form a complete basis for the subspace of band-limited signals, $f(\hat{\boldsymbol{x}})$ can be represented in the Slepian basis using (26). We plot the spherical harmonic and Slepian coefficients of $f(\hat{\boldsymbol{x}})$ in descending order of their magnitude in Fig. 7 where, as expected, the Slepian coefficients decay more quickly than the spherical harmonic coefficients. The spatially concentrated signal has a sparse representation in the Slepian basis, it can be represented accurately using $N = 546$

[2] http://www.sdss.org/dr7/

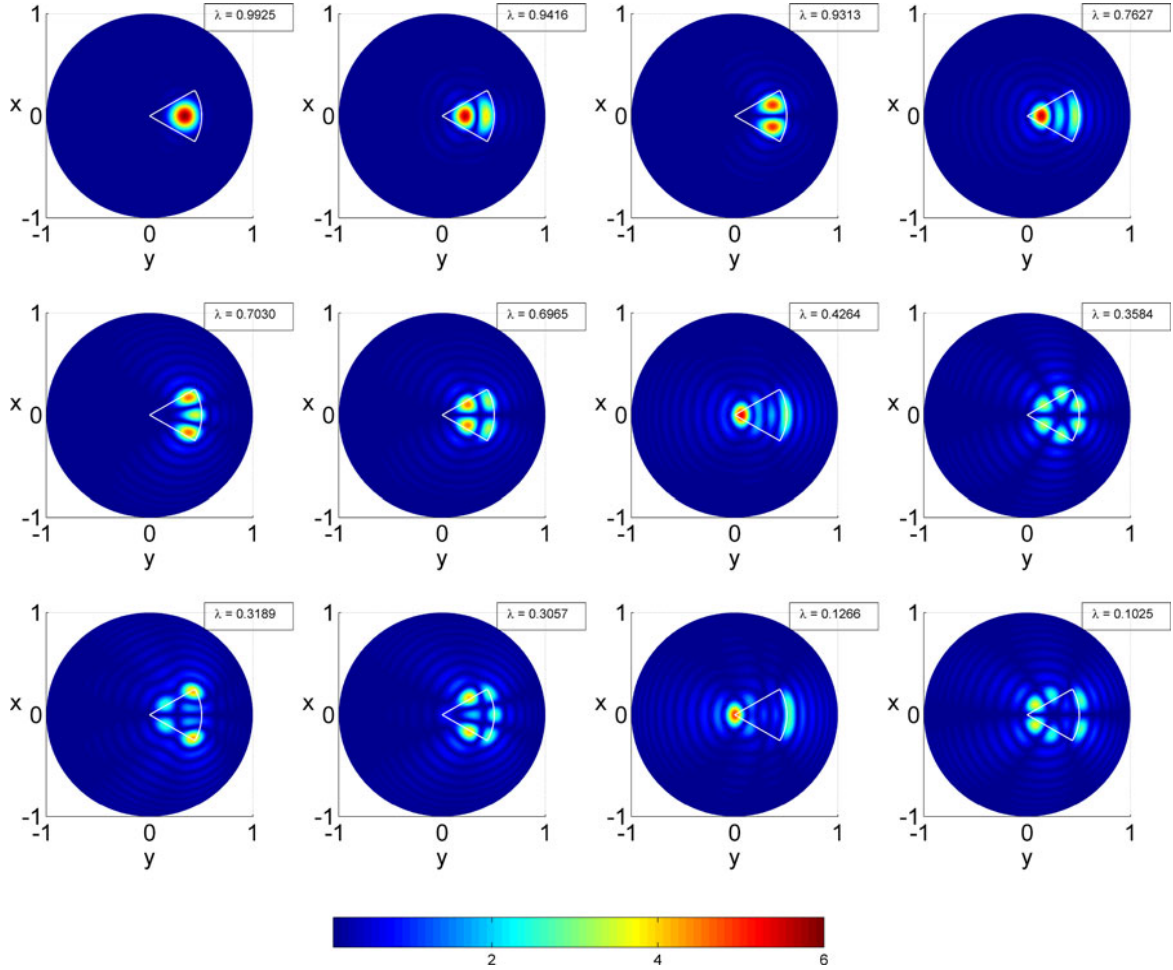

Fig. 3. Magnitude of Slepian functions $|h_\alpha(\hat{\boldsymbol{x}})|$, $\alpha = 1, 2, \ldots, 12$ concentrated in $\tilde{R}$ in Example A, shown with their corresponding eigenvalue $\lambda_\alpha$, with band-limit $L = 25$ ($N = 7$). The ordering of concentration is left to right, top to bottom.

Slepian coefficients, shown by the red dashed line in Fig. 7, rather than $L^2 = 2500$ spherical harmonic coefficients. Fig. 6(b) shows the signal reconstructed by expansion in the truncated Slepian basis using (29) with $J = N = 546$ Slepian functions. The energy ratio given in (30), is $Q(N = 546) = 99.7\%$ quantifying that the approximation is sufficiently accurate with the spatial region of interest.

## IV. FAST COMPUTATION OF **K**

In this section, we devise a formulation for the fast computation of the matrix **K** using the analytic expression presented in Theorem 1 in Section III-A. By exploiting the symmetry relations exhibited by spherical harmonics, we also propose an implementation to reduce the computation time.

The proposed analytic expression to compute the matrix elements $E_{\ell m, pq}$, given in Theorem 1 can be rewritten as

$$K_{\ell m, pq} = S(q-m) \sum_{m'=-\ell}^{\ell} \sum_{q'=-p}^{p} F^\ell_{m',m} F^p_{q',q} Q(m'+q'). \quad (32)$$

For a given band-limit $L$, we need to compute $L^4$ elements of the matrix **K**, that is, $K_{\ell m, pq}$ is required to be computed for each $\ell, p < L$, $|m| \leq \ell$ and $|q| \leq p$. Naively, the computation complexity to compute $L^4$ elements of **K** is $O(L^6)$ as the computation of $K_{\ell m, pq}$, using (32), requires the evaluation of two summations, each with the maximum order of $L$. Using separation of variables, $K_{\ell m, pq}$ in (32) can be reformulated as

$$K_{\ell m, pq} = S(q-m) B_{\ell m, pq}, \quad (33)$$

with

$$B_{\ell m, pq} \triangleq \int_{\theta_1}^{\theta_2} Y_\ell^m(\theta, 0) \overline{Y_p^q(\theta, 0)} \sin\theta d\theta$$

$$= \sum_{m'=-\ell}^{\ell} F^\ell_{m',m} C^p_{m',q}, \quad (34)$$

where

$$C^p_{m',q} = \sum_{q'=-p}^{p} F^p_{q',q} Q(m'+q'). \quad (35)$$

We note that each $\ell, m, p, q, m', q'$ has the maximum order or degree of $L$. Using (35), $C^p_{m',q}$ can be computed for all $p < L$, $|q| < p$ and $|m'| < L$ in $O(L^4)$ time. Once we have $C^p_{m',q}$, $B_{\ell m, pq}$ can be computed using (34) for all $\ell, p < L$, $|m| \leq \ell$, $|q| \leq p$, that is, a total of $L^4$ elements, with computational complexity $O(L^5)$. $E_{\ell m, pq}$ is then computed in $O(L^4)$ using (33), resulting in the overall complexity of $O(L^5)$, compared to the naive scaling of $O(L^6)$.

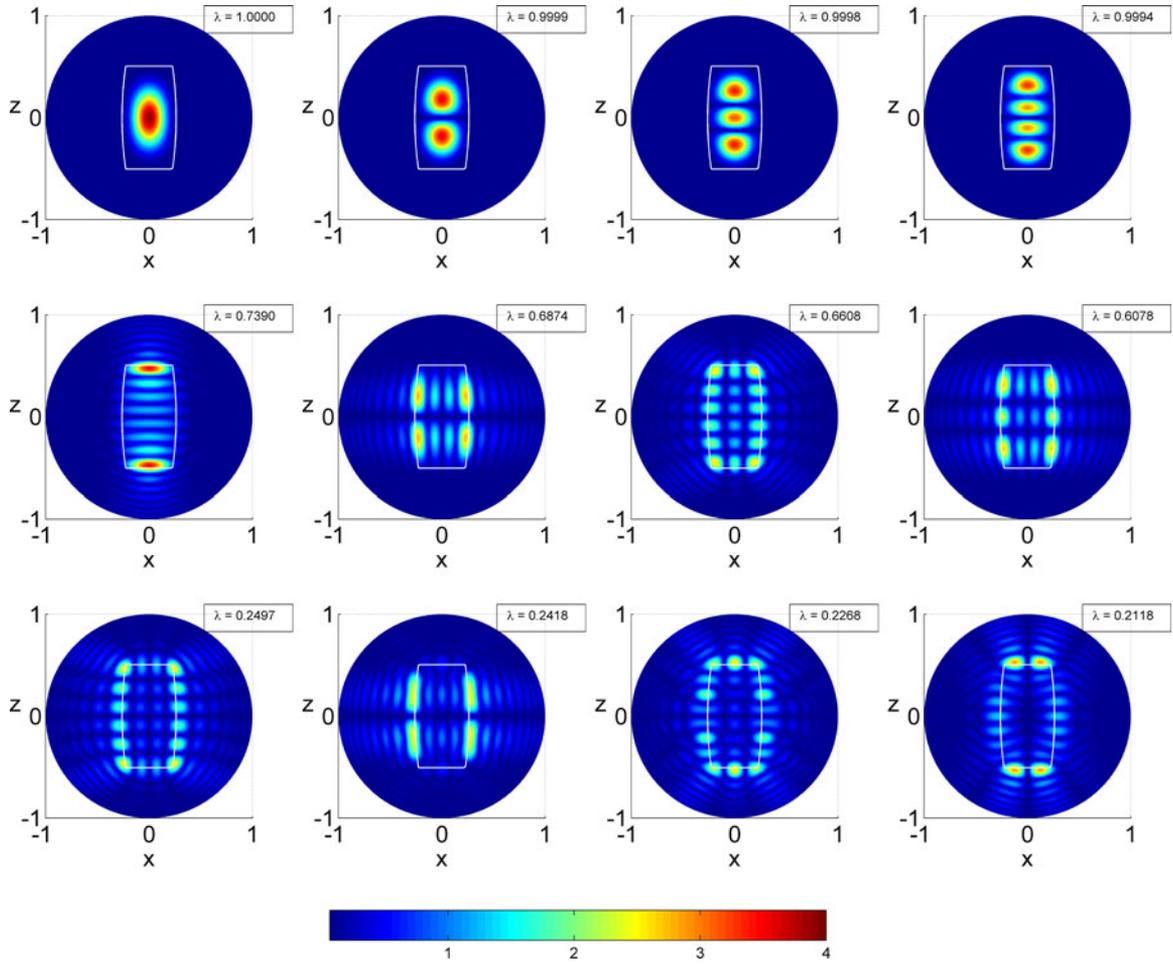

Fig. 4. Magnitude of Slepian functions $|h_\alpha(\hat{\boldsymbol{x}})|$, $\alpha = 1, 2, 3, 4, 21, 22, 23, 24, 31, 32, 33$ and $34$ concentrated in $\tilde{R}$ in Example B, shown with their corresponding eigenvalue $\lambda_\alpha$, with band-limit $L = 25$ ($N = 26$). The ordering of concentration is left to right, top to bottom.

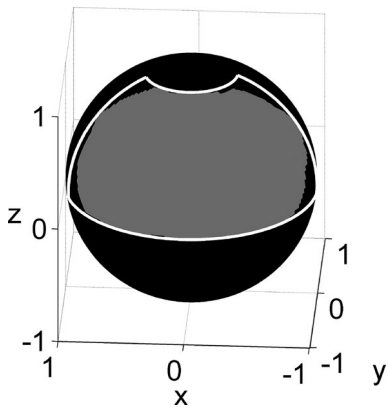

Fig. 5. The limited colatitude-longitude region (outline shown in white) approximating the SDSS DR7 quasar mask on the sphere (shown in grey).

*Remark 3 (On the use of FFT for the computation of (35)):* For each $p < L$ and $|q| < p$, the complexity to compute $C^p_{m',q}$ for all $m' < L$ is $O(L^2)$. Noting that the summation involved in the computation of $C^p_{m',q}$ using (35), is in the form of a discrete convolution, which offers an opportunity to employ fast Fourier transforms (FFT) to carry out this step in $O(L \log_2 L)$ as

$$C^p_{m',q} = \mathcal{F}^{-1}\Big(\mathcal{F}(F^p_{-q',q})\mathcal{F}\big(Q(q')\big)\Big), \qquad (36)$$

where $\mathcal{F}$ and $\mathcal{F}^{-1}$ denote FFT and inverse FFT respectively. The use of FFT reduces the complexity of the computation of $C^p_{m',q}$ for all $p < L$, $|q| < p$ and $|m'| < L$, $q$ from $O(L^4)$ to $O(L^3 \log_2 L)$.

We note that the use of FFT improves the computational complexity to compute $C^p_{m',q}$ and consequently reduces the overall computation time; however, does not alter the overall complexity $O(L^5)$.

*Remark 4 (Computation for Arbitrary Region of Interest):* Naively the computational complexity of computing $\mathbf{K}$ for an arbitrary region given by (31) appears to be $O(ML^6)$. However using separation of variables and computing in matrix form reduces the computational complexity to $O(ML^5)$, which is intuitively the number of regions $M$ times the computational complexity of computing $\mathbf{K}$ for a single limited colatitude-longitude region $\tilde{R}$.

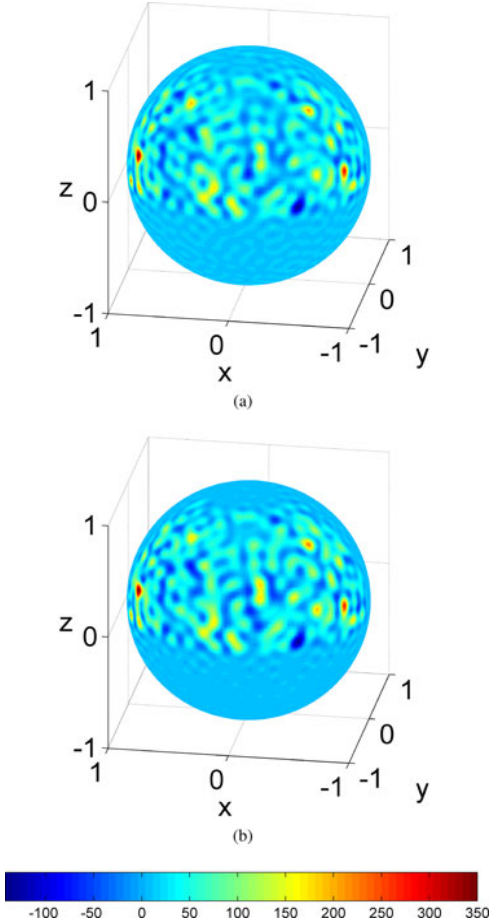

Fig. 6. (a) Simulated dark matter test signal on the sphere $f(\hat{\boldsymbol{x}})$ band-limited at $L = 50$ and spatially concentrated within limited colatitude-longitude region shown in Fig. 5 (b) test signal reconstructed by expansion in the truncated Slepian basis using $N = 546$ Slepian functions.

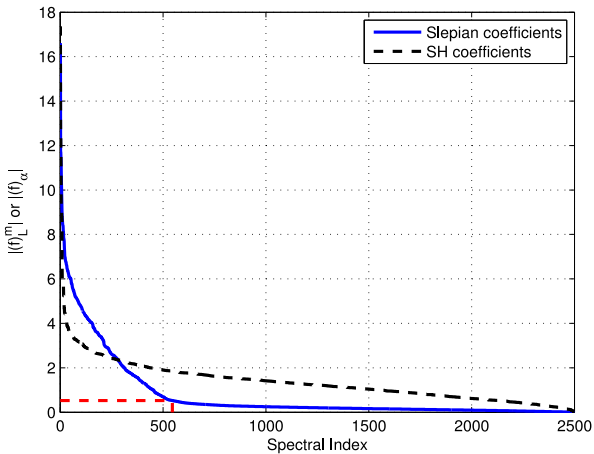

Fig. 7. Spectral decay of the spherical harmonic (SH) (black dashed line) and Slepian (blue solid line) coefficients of the band-limited spatially concentrated test signal $f(\hat{\boldsymbol{x}})$ shown in Fig. 6. The magnitude of the spherical harmonic and Slepian coefficients are plotted in descending order of magnitude. The sum of eigenvalues $N = 546$, given by (25), is shown by the dashed red line.

### A. Computation Time Reduction

The overall computation time can be further reduced by exploiting the symmetry relations exhibited by spherical harmonics. As mentioned earlier, the matrix is Hermitian symmetric, that is,

$$K_{pq,\ell m} = \overline{K_{\ell m, pq}}. \quad (37)$$

Consequently, we are only required to compute half of the off-diagonal elements of the matrix $\mathbf{K}$. Furthermore, we note that the computation of $K_{\ell m,pq}$ from $B_{\ell m,pq}$ only requires scaling by a factor $S(q-m)$ as given by (33) and is therefore can be carried out quickly in $O(L^4)$ for all $\ell, m, p$ and $q$. For the computation of $B_{\ell m,pq}$, we also use the following symmetry relations

$$B_{\ell m,pq} = (-1)^m B_{\ell(-m),pq} = (-1)^{m+q} B_{\ell(-m),p(-q)}, \quad (38)$$

to speed up the computation. These symmetry relations stem from the following symmetry relation of spherical harmonics[3] (and associated Legendre polynomials) [35]

$$Y_\ell^m(\theta, 0) = (-1)^m Y_\ell^{-m}(\theta, 0). \quad (39)$$

For each $\ell < L$ and $p < L$, we need to compute $B_{\ell m,pq}$ for $(2\ell+1) \times (2p+1)$ times, that is, for each $|m| \leq \ell$ and $|q| \leq p$. Instead, we compute one fourth of these $(2\ell+1) \times (2p+1)$ elements, that is for each $m \leq \ell$ and $q \leq p$. The remaining elements can be computed by exploiting the symmetry relations in (38). Summarizing, the use of symmetry relations in (37) and (38) reduces the computation time, approximately, by a factor of 8.

### B. Computation of Wigner-$d$ Functions

In the computation of $\mathbf{K}$ using (33)-(35), we still need to address the computation of $F_{m',m}^\ell$, given in (17), which, in turn, requires the computation of Wigner-$d$ functions $\Delta_{m,n}^\ell$ for all $\ell < L$ and $|m|, |m'| \leq \ell$. Let $\Delta_\ell$ denote the matrix of size $(2\ell+1) \times (2\ell+1)$ with entries $\Delta_{m,n}^\ell$ for $|n|, |m| \leq \ell$. The matrix $\Delta_\ell$ can be computed for each $\ell = 1, 2, \ldots, L-1$ using the relation given in [44] that recursively computes $\Delta_\ell$ from $\Delta_{\ell-1}$. Alternative to the recursion relation proposed in Trapani and Navaza [44], we note that the recursion relation proposed by Risbo [38] can also be employed for the computation of $\Delta_{m,n}^\ell$. It must be noted that these recursions are stable up to very large band-limits.

Computation of $F_{m',m}^\ell$ for all $\ell < L$, $|m'| \leq \ell$ and $|m| \leq \ell$ has computational complexity $O(L^3)$, and therefore does not change the overall complexity of the algorithm. In addition, as $F_{m',m}^{\ell}$ does not depend on the limited colatitude-longitude region parameters it can be precomputed. Precomputation of $F_{m',m}^{\ell}$ requires $O(L^3)$ storage which is less than the that required to store $\mathbf{K}$ which is $O(L^4)$.

[3] The symmetry relation in (39) follows from the adopted definition of spherical harmonics with Condon-Shortley phase included, due to which we have the preceding factor $(-1)^m$ on the right hand side of (39). For alternative definitions of spherical harmonics that do not include this phase factor, we note that the modified symmetry relations can be formulated and exploited to speed up the computation.

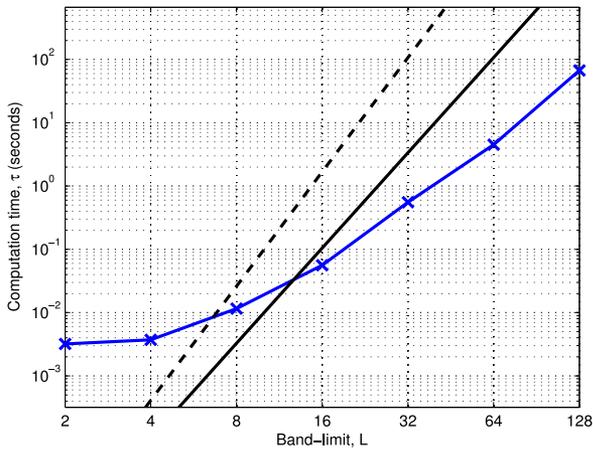

Fig. 8. Computation time $\tau$ in seconds to calculate the matrix $\mathbf{K}$ for a limited colatitude-longitude region (solid blue line) using the algorithm presented in Section IV for band-limits $L = 2^n$, $n = 1, 2, \ldots 7$. The computation time scales as $O(L^5)$ (solid black line) rather than $O(L^6)$ (dashed black line).

### C. Computational Time Analysis

We calculated the computation time in seconds, denoted by $\tau$, to carry out the algorithm presented in Section IV for calculation of the matrix $\mathbf{K}$ for a limited colatitude-longitude region. Fig. 8 shows the computation time verse band-limit $L$ to calculate $\mathbf{K}$ averaged over 100 iterations for $L = 2^n, \forall n \in [1, 7]$. The computation is performed using MATLAB running on a machine equipped with 3.4 GHz Intel Core i7 processor and 8 GB of RAM. In Fig. 8 it can be seen that the algorithm scales closer to $O(L^5)$ than $O(L^6)$ as was expected from the computational complexity analysis in Section IV.

Our proposed algorithm has an inherently parallel structure, therefore parallel computing methods could be used to further reduce the time required to compute the limited colatitude-longitude region Slepian functions.

## V. CONCLUSION

We have developed an analytical formulation for the Slepian spatial-spectral concentration problem on the sphere for a limited colatitude-longitude spatial region on the sphere that enables accurate computation of the Slepian functions and eigenvalues for this region. We also extended this analytical formulation for an arbitrary region on the sphere comprised of a union of rotated limited colatitude-longitude subregions. In addition, we have developed a computationally efficient algorithm for implementation of the proposed analytical formulation. We perform computational complexity analysis of our algorithm and use examples to illustrate the use of our algorithm in applications. Future work includes applying our algorithm to applications where limited colatitude-longitude regions on the sphere occur such as modeling the direction of arrival in communications, direction of sound projected from a rectangular speaker in acoustics and optics, and the modeling of cosmic microwave background radiation in astrophysics.

## ACKNOWLEDGMENT

The authors would like to thank Dr. B. Leistedt and Dr. J. D. McEwen for providing the Horizon Simulation dataset. And also would like to thank F. J. Simons for discussions, and the Department of Geosciences at Princeton University for their hospitality during A. P. Bates' visit in the Fall of 2015.

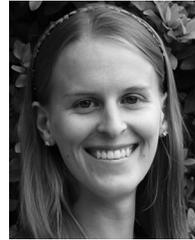

**Alice P. Bates** (S'11–M'16) received the B.E. (first class Hons.) degree in electrical engineering from the University of Auckland, Auckland, New Zealand, in 2013, and the Ph.D. degree in engineering from the Australian National University (ANU), Canberra, ACT, Australia, in September 2016.

She is currently a Research Fellow at the Research School of Engineering, ANU. Her research interests include the application-driven development of signal processing techniques for the collection and processing of signals with spherical geometry. She received the Senior Scholar Award for first in her cohort during her undergraduate studies. She also received an Australian Postgraduate Award for the duration of her Ph.D.

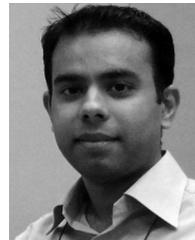

**Zubair Khalid** (S'10–M'13) received the B.Sc. (first class Hons.) degree in electrical engineering from the University of Engineering and Technology, Lahore, Pakistan, in 2008, and the Ph.D. degree in engineering from the Australian National University (ANU), Canberra, ACT, Australia, in August 2013.

He is currently an Assistant Professor in the Department of Electrical Engineering, University of Engineering and Technology, Lahore, Pakistan. He was a Research Fellow in the Research School of Engineering, ANU. His research interests are in the area of signal processing and wireless communications, including the development of novel signal processing techniques for signals on the sphere and the application of stochastic geometry in wireless *ad-hoc* networks. He received the University Gold Medal and Industry Gold Medals from Siemens and Nespak for his overall outstanding performance in electrical engineering during his undergraduate studies and an Endeavour International Postgraduate Award for his Ph.D. studies.

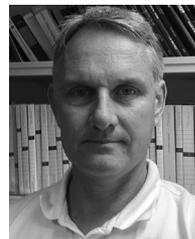

**Rodney A. Kennedy** (S'86–M'88–SM'01–F'05) received the B.E. degree (first class honours and university medal) from the University of New South Wales, Sydney, NSW, Australia, the M.E. degree from the University of Newcastle, Callaghan, NSW, and the Ph.D. degree from the Australian National University, Canberra, ACT, Australia.

Since 2000, he has been a Professor in engineering at the Australian National University, Canberra. He has coauthored more than 300 refereed journal or conference papers and a book *Hilbert Space Methods in Signal Processing* (Cambridge Univ. Press, 2013). He has been a Chief Investigator in a number of Australian Research Council Discovery and Linkage Projects. His research interests include digital signal processing, digital and wireless communications, and acoustical signal processing.